\documentclass[letterpaper, 10 pt, conference]{ieeeconf}  
\usepackage{float}

\IEEEoverridecommandlockouts                              

\usepackage{xcolor}
\usepackage{soul}
\usepackage{graphicx}
\usepackage{comment}
\usepackage{amsmath, amssymb}
\usepackage{color}

\title{\LARGE \bf
A Framework for Ethical Decision-Making in Automated Vehicles through Human Reasons-based Supervision
}

\author{Lucas Elbert Suryana$^{1,3}$, Saeed Rahmani$^{1}$, Simeon C. Calvert$^{1,3}$, Arkady Zgonnikov$^{2,3}$, Bart van Arem$^{1}$
\thanks{$^{1}$Department of Transport and Planning, $^{2}$Department of Cognitive Robotics, $^{3}$Centre for Meaningful Human Control, Delft University of Technology, 2628 CN Delft, The Netherlands. {\tt\small l.e.Suryana@tudelft.nl; s.rahmani@tudelft.nl; s.c.calvert@tudelft.nl; b.vanarem@tudelft.nl; a.zgonnikov@tudelft.nl.}}%
}

\begin{document}

\maketitle
\thispagestyle{empty}
\pagestyle{empty}

\begin{abstract}
Ethical dilemmas are a common challenge in everyday driving, requiring human drivers to balance competing priorities such as safety, efficiency, and rule compliance. However, much of the existing research in automated vehicles (AVs) has focused on high-stakes "trolley problems," which involve extreme and rare situations. Such scenarios, though rich in ethical implications, are rarely applicable in real-world AV decision-making. In practice, when AVs confront everyday ethical dilemmas, they often appear to prioritise strict adherence to traffic rules. By contrast, human drivers may bend the rules in context-specific situations, using judgement informed by practical concerns such as safety and efficiency. According to the concept of meaningful human control, AVs should respond to human reasons, including those of drivers, vulnerable road users, and policymakers. This work introduces a novel human reasons-based supervision framework that detects when AV behaviour misaligns with expected human reasons to trigger trajectory reconsideration. The framework integrates with motion planning and control systems to support real-time adaptation, enabling decisions that better reflect safety, efficiency, and regulatory considerations. Simulation results demonstrate that this approach could help AVs respond more effectively to ethical challenges in dynamic driving environments by prompting replanning when the current trajectory fails to align with human reasons. These findings suggest that our approach offers a path toward more adaptable, human-centered decision-making in AVs.
\end{abstract}


 


\section{Introduction}\label{introduction}
Addressing the ethical complexities that emerge in daily driving contexts remains essential for social acceptance of automated vehicles (AVs). Despite their promised advantages in safety improvements and transportation access \cite{geisslinger2021autonomous}, the widespread adoption of these systems hinges on their capacity to reflect human ethical judgment, particularly when confronting morally ambiguous situations where multiple values compete \cite{lin2016ethics, millar2017ethics}---situations commonly referred to as ethical dilemmas. Examples include deciding whether to briefly occupy the opposite lane to safely overtake cyclists \cite{TeslaShort2025}, or speeding up temporarily to avoid unsafe situations. This behaviour reveals a critical gap in AV decision-making: the necessity of designing AVs capable of dynamically balancing multiple considerations, such as safety, efficiency, regulatory compliance, and contextual appropriateness, in real-time, rather than relying solely on predefined regulations.

Current approaches show limitations when it comes to handling everyday ethical dilemmas in automated driving. Much of the research has focused on extreme scenarios, such as the well-known ``trolley problem'' \cite{bonnefon2019trolley}. While philosophically significant, trolley problems rarely happen in daily driving, and despite the practical importance of everyday dilemmas, they often receive less attention \cite{geisslinger2021autonomous, himmelreich2018never}. \cite{lin2016ethics} emphasises that everyday ethical decisions in automated driving extend far beyond these extreme scenarios, requiring nuanced contextual use of reasons, something current systems lack. Similarly, \cite{nyholm2016ethics} argue that framing AV ethics as simplified trolley problems fails to capture the probabilistic nature and dynamic complexity of real-world driving situations. 

Although the ethical dimensions of ``mundane'' scenarios may seem straightforward for human drivers, they require context-aware judgment that comes naturally to human but poses challange for AVs. These judgments must balance multiple ethically relevant considerations, such as safety, efficiency, and social norms. This adaptability represents a challenge for AV systems designed with traditional motion planning algorithms, which primarily optimise for trajectory smoothness and collision avoidance without explicitly integrating ethical considerations. Building on this understanding, recent ethical frameworks propose more holistic approaches that better align with human moral intuitions. \cite{cecchini2024aligning} and \cite{henschke2020trust} collectively emphasise that effective AV ethics must integrate moral principles such as deontological ethics, virtue ethics, and consequentialist considerations while maintaining transparency in decision-making. 

However, integrating these ethical principles into AV decision-making remains challenging. While prior works have focused on embedding such principles into control and motion algorithms \cite{geisslinger2021autonomous, thornton2016incorporating, geisslinger2023ethical}, current approaches fail to make explicit when these principles are in conflict. Recognising these conflicts is essential for enabling transparent decisions and for adjusting AV behaviour to better reflect the ethical principles the system is intended to uphold.

The concept of meaningful human control (MHC) \cite{mecacci2020meaningful, santoni2018meaningful} offers a promising conceptual bridge for addressing the challenge of making explicit which moral principles are in conflict. MHC asserts that humans should ultimately be responsible for every decision made by automated systems. \cite{santoni2018meaningful} laid the groundwork for achieving MHC. One of the required conditions is the tracking condition, which requires automated systems to respond to the reasons of relevant humans. In the remainder of this paper, we refer to these relevant humans -- such as drivers, passengers, pedestrians, and policymakers -- as stakeholders. According to \cite{mecacci2020meaningful}, these reasons can be understood as moral values or principles that are reflected in human driving plans and intentions—such as ensuring safety and comfort for both themselves and others, driving efficiently, and complying with traffic regulations. From this perspective, if an AV is designed to uphold certain moral principles, the tracking condition provides a clear expectation that its behaviour should reflect corresponding human plans and intentions.

To operationalise this concept and address the challenges of handling ethical dilemmas and making moral principles explicit in AV decision-making, we propose a novel human reasons-based supervision framework that enables AVs to evaluate if their behaviour aligns with the reasons of diverse stakeholders. By grounding this framework in the tracking condition of meaningful human control, we aim to support AV decision-making in ethically challenging everyday scenarios that require balancing multiple, sometimes conflicting, values.

Specifically, the primary contribution of this paper is a modular human reasons-based supervision framework that enables AVs to make ethically nuanced decisions in routine yet ethically challenging scenarios. The framework continuously evaluates how well the AV's behaviour aligns with human reasons and triggers replanning when a misalignment is detected. The paper contributes:
\begin{enumerate}
\item We developed a detection mechanism that uses stakeholder reason scores and predefined thresholds to identify when AV behaviour misaligns with human reasons;
\item We integrated the human reasons-based supervision framework into an AV control architecture, including a mechanism for triggering replanning when reason scores fall below predefined thresholds;
\item We enabled explainability by using reason scores as interpretable indicators of why behaviour changes are recommended in routine, ethically challenging situations.
\end{enumerate}

This work advances the discourse on AV decision-making in ethically challenging transportation scenarios by bridging the gap between moral principles and practical AV decision-making, ultimately supporting the development of socially acceptable automated mobility solutions that align with human reasons across diverse everyday scenarios.

The remainder of this paper is organised as follows: Section \ref{sec:methodology} presents the detailed methodology and system architecture, including the mathematical human reasons and its integration into a motion planning framework. Section \ref{sec:experiment} describes the experimental setup and simulation environment. Section \ref{sec:results} and \ref{sec:discussion} present and discuss the simulation results and the impact of ethical supervision on vehicle behaviour. Finally, Section \ref{sec:conclusion} concludes the paper and outlines directions for future research.

\section{Methodology}\label{sec:methodology}
\subsection{Problem Formulation}
We formalise the automated vehicle navigation problem in scenarios involving ethical decision-making. While the framework is generic and applicable to a wide range of situations, for demonstration purposes, we consider a scenario including the interaction of an automated vehicle with a vulnerable road user (VRU). In this scenario, an automated vehicle navigating a bidirectional road faces an ethical dilemma during overtaking manoeuvres. To maintain efficient travel, the vehicle must either follow the VRU with a very low speed, which is not desirable for the vehicle's passenger, or overtake a slower-moving VRU, which may require temporarily entering the oncoming lane or reducing the safety buffer with the cyclist. This manoeuvre challenges forces a trade-off between strict compliance, user safety, and travel efficiency.

For the problem formulation, we consider the automated vehicle operating in state space $\mathcal{X} \subset \mathbb{R}^n$ with state vector $\mathbf{x_t} = [\mathbf{p_t}, v_t, \theta_t, \omega_t]^{T}$, where $\mathbf{p_t} = [p_x, p_y]^{T}$ represents position, $v_t$ denotes velocity, $\theta_t$ is heading angle, and $\omega_t$ is rotational velocity. The control space $\mathcal{U} \subset \mathbb{R}^m$ consists of $\mathbf{u_t} = [a_t, \delta_t]^{T}$, representing acceleration and steering angle.

Our multi-agent ethical framework defines stakeholders $\mathcal{S} = \{s_1, s_2, ..., s_k\}$ with reason functions $R_{s_i}: \mathcal{X} \times \mathcal{U} \rightarrow [0,1]$ quantifying alignment with each stakeholder's perspective. This formulation means that each function $R_{s_i}$ evaluates the vehicle's state and control actions to produce a score between 0 and 1, reflecting how well the AV's behaviour satisfies the ethical priorities of the respective stakeholder. For the designed scenario, we identify three key stakeholders: road policymakers ($s_{policy}$), vulnerable road user ($s_{VRU}$), and drivers ($s_{driver}$).

The navigation problem for the automated vehicle is formulated as:
\begin{equation}
\begin{split}
\min_{u_0,...,u_{T-1}} &\sum_{t=0}^{T-1} \mathcal{J}(\mathbf{x_t}, \mathbf{u_t}) \\
\text{s.t.}\quad &\mathbf{x_{t+1}} = f(\mathbf{x_t}, \mathbf{u_t}) \\
&\mathbf{x_t} \in \mathcal{X}_{safe} \\
&\mathbf{u_t} \in \mathcal{U} \\
&R_{s_i}(\mathbf{x_t}, \mathbf{u_t}) \geq \tau_{s_i}, \forall s_i \in \mathcal{S}
\end{split}
\end{equation}

Here, $\mathcal{J}(\mathbf{x_t}, \mathbf{u_t})$ represents the cost function to be minimised over the control horizon $T$, evaluating the performance of the AV’s state and control inputs at each time step $t$. The function $f(\mathbf{x_t}, \mathbf{u_t})$ denotes the system dynamics model that predicts the next state $\mathbf{x_{t+1}}$ based on the current state $\mathbf{x_t}$ and control input $\mathbf{u_t}$. The set $\mathcal{X}_{safe} \subset \mathcal{X}$ defines the safe region of the state space where the vehicle must operate to avoid collisions and other hazards. Finally, for each stakeholder $s_i$, $\tau{s_i}$ denotes the threshold value specifying the minimum acceptable reason score that the AV’s behaviour must meet.

\subsection{Framework Architecture}
Our approach implements a multi-component and hierarchical framework with three main components:
\begin{enumerate}
    \item \textit{Global Motion Planning}: Responsible for finding a reference trajectory for the vehicle to be followed. It uses A* search with motion primitives to generate feasible reference paths from the current state of the vehicle to the goal location.
    \item \textit{Model Predictive Control}: Optimises vehicle trajectory when following the reference path. It ensures kinodynamic feasibility and satisfying soft and hard constrained defined, such as safety, efficiency, and comfort.
    \item \textit{Human Reasons-based Supervision Framework}: Evaluates the planned actions against ethical criteria and triggers replanning when necessary if the criteria are not met.
\end{enumerate}

\begin{figure*}
    \centering
    \includegraphics[width=0.8\linewidth]{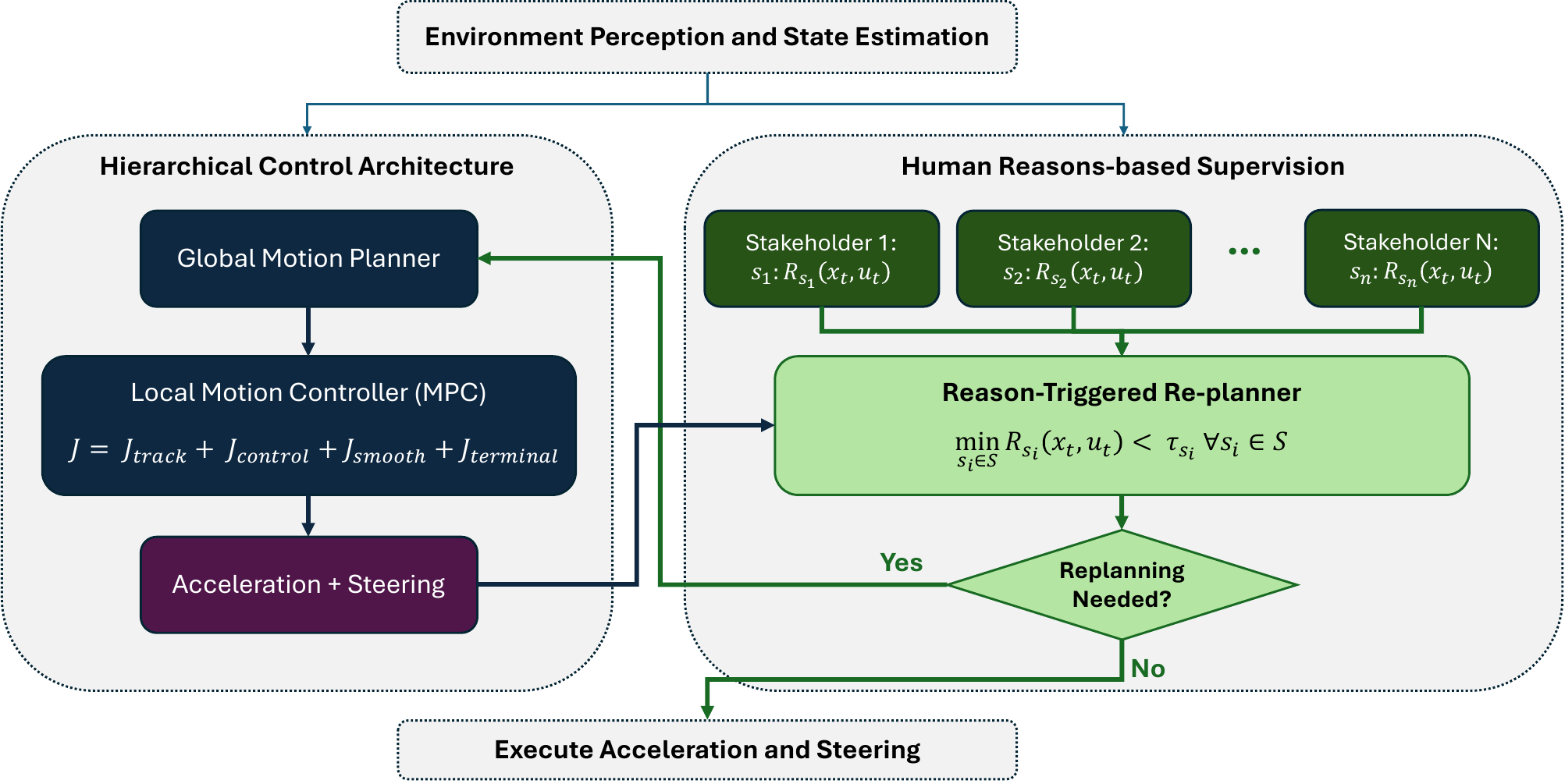}
    \caption{Hierarchical control architecture with human reasons-based supervision for ethical AV decision-making}
    \label{fig:framework}
\end{figure*}

These elements are depicted in Fig. \ref{fig:framework}. The key innovation in our approach is the definition and integration of a human reasons-based supervision framework as a mechanism for triggering replanning, ensuring that the vehicle's behaviour satisfies ethical constraints derived from multiple stakeholders' perspectives. Therefore, we begin by detailing the components of the framework.

\subsection{Human Reasons-based Supervision Framework}
Our approach to developing a framework that supervises the alignment between AV behaviour and human reasons builds on the qualitative evaluation steps for the tracking condition outlined by \cite{suryana2025meaningful}. These steps involve defining the relevant stakeholders and articulating their reasons, as well as specifying the features of the AV system that govern its behaviour. While the original approach remains qualitative, our work extends it by developing a quantitative framework. Specifically, after identifying the stakeholders, we formalise their reasons mathematically. This process is described in this section, while the features of the AV system that govern its behaviour are presented in Section~\ref{sectionD}.

\subsubsection{Identification of Stakeholders and Reason Models}
To effectively integrate human ethical considerations into AV decision-making, it is essential to identify the key stakeholders involved and define how system's behaviour influence their alignment with stakeholders' reasons. Accordingly, we model three primary stakeholders in the designed scenario:

\begin{itemize}
    \item \textit{Road Policymaker}: Represents regulatory authorities whose reason is to ensure overall road safety through regulatory compliance.
    \item \textit{Vulnerable Road User}: Represents vulnerable road users whose reason is to commute with safety and comfort.
    \item \textit{Driver}: Represents the vehicle occupant's motivation for efficiency and arriving at the destination as fast as possible.
\end{itemize}

To operationalise stakeholder perspectives within our framework, we establish specific reason models for each. For each stakeholder, we define a reason model that quantifies their satisfaction with the vehicle's behaviour on a scale from 0 to 1, where 1 indicates full satisfaction and 0 indicates complete dissatisfaction.

\subsubsection{Mathematical Representation of Reason Models}

A key contribution of our framework is the operationalisation of abstract human reasons using empirically supported, measurable parameters. While previous work has discussed human reasons conceptually \cite{suryana2025meaningful} or proposed initial variables \cite{calvert2020conceptual}, these efforts have not established an empirically grounded mapping to real-world driving variables. We address this gap by introducing a reason model grounded in human factors research. 

To translate stakeholder's reason into a computationally viable form, we adopt a set of piecewise exponential functions that model how stakeholder satisfaction declines when specific behavioural threshold are crossed. These modelling choice are grounded based on both computational simplicity and their ability to approximate human reasons. As demonstrated by \cite{tversky1992advances}, individuals tend to overweight low probabilities and underweight high probabilities, implying a rapid shift in perceived risk once certain thresholds are crossed.  Thus, the exponential function is chosen because it can well capture the rapid change in perceived acceptability. Nevertheless, the proposed equations serve as a representative model that can be adjusted via thresholds and scaling constants to suit various scenarios.

To ensure realism, each stakeholder's reason is modelled using scenario-specific variables informed by human factors research. Details of the experimental case appear in Section~\ref{sec:experiment}. For example, the cyclist’s reason—related to comfort and perceived safety—is represented using lateral distance and tailgating time, based on findings from \cite{rsa2018mpd, oskina2023safety}. Here, comfort refers to the cyclist’s subjective experience of emotional and physical ease during interaction with the vehicle. Empirical studies show that insufficient lateral clearance and prolonged close following increase stress and perceived risk, justifying the use of these variables as proxies.

The same vehicle behaviour--prolonged following--also impact driver's reason, which relates to driving efficiency. This is modelled as perceived impatience, based on time and distance the AV follows the cyclist at close range.  As shown in \cite{lee2010measuring}, such conditions could lead to frustration under time pressure. Meanwhile, the policymaker’s reason—regulatory compliance—is satisfied when the AV stays within its designated lane. The reason's score decreases as the AV crosses into the opposite lane to overtake the cyclist. This reflects findings from \cite{suryana2025principles}, where experts preferred the AV to overtake gently and return to its lane promptly, indicating that in such situations, any lane violation should be considered minimal.

\paragraph{Policymaker's Reason}  
The policymaker's reason score quantifies regulatory compliance -- specifically, adherence to lane regulations in this scenario:

\begin{equation}\label{eq:policy_reason}
R_{\text{policymaker}} = 
\begin{cases} 
1, & \text{if } d_{\text{veh}} > 0, \\
e^{k_1 \cdot d_{\text{veh}}}, & \text{otherwise},
\end{cases}
\end{equation}
\normalsize

where \(d_{\text{veh}}\) is the lateral displacement of the ego vehicle from the center line (positive when on the correct side of the road, negative when in the oncoming lane); \(k_1\) is a scaling constant.

\paragraph{VRU's Reason}  
The VRU's reasons score is decomposed into safety assurance and comfort preservation.

- \textit{Safety Assurance:}
\begin{equation}
R_{sa}(t) = 
\begin{cases} 
1, & \text{if } d_{\text{veh-vru}} > d_{\text{th,vru}}, \\
\frac{1}{e^{k_{2}\,(d_{\text{veh-vru}} - d_{\text{th,vru}})}}, & \text{otherwise},
\end{cases}
\end{equation}
\normalsize

where \(d_{\text{veh-vru}}\) denotes the distance between the vehicle and the VRU; \(d_{\text{th,vru}}\) is the perceived safe distance threshold; \(k_{2}\) is a scaling constant.

- \textit{Comfort Preservation:}
\begin{equation}
R_{cp}(t) = 
\begin{cases} 
1, & \text{if } t_{\text{close,vru}} < t_{\text{th,vru}} \text{ or } \\ & d_{\text{veh-vru}} > d_{\text{th,vru}}, \\
\frac{1}{e^{k_{3}\,(t_{\text{follow}} - t_{\text{th,vru}})}}, & \text{otherwise}.
\end{cases}
\end{equation}
\normalsize

where \(t_{\text{close,vru}}\) is the cumulative time during which \(d_{\text{veh-vru}} < d_{\text{th,vru}}\); \(t_{\text{th,vru}}\) is the maximum tolerable time for the cyclist to be followed too closely; \(k_{3}\) is a scaling constant.

The overall VRU's reason score is then given by:
\begin{equation}\label{eq:vru_reason}
R_{\text{VRU}}(t) = R_{sa}(t) \cdot R_{cp}(t).
\end{equation}
where \(R_{\text{VRU}}(t)\) combines the safety and comfort components.

\paragraph{Driver's Reason}  
The driver's reason score is defined as:
\begin{equation}
R_{\text{driver}}(t) = 
\begin{cases} 
1, & \text{if } t_{\text{behind,driver}} < t_{\text{th,driver}} \text{  or }\\
    &  \text{    } d_{\text{veh-vru}} > d_{\text{th,driver}}, \\
\frac{1}{e^{k_4 (t_{\text{behind,driver}} - t_{\text{th,driver}})}}, & \text{otherwise},
\end{cases}
\label{eq:vehicle_reason}
\end{equation}

where \(t_{\text{behind,driver}}\) is the cumulative time during which \(d_{\text{veh-vru}} < d_{\text{th,driver}}\); \(t_{\text{th,driver}}\) is the time threshold for close following that the driver considers acceptable; \(d_{\text{veh-vru}}\) is the distance between the vehicle and the VRU; \(d_{\text{th,driver}}\) is the distance threshold below which the driver considers the AV to be following too closely, leading to perceived inefficiency; and \(k_{4}\) is a scaling constant. Note that the scores of \(k_{1},k_{2},k_{3},\) and \(k_{4} = 0.2\) in our experiment can be adjusted depending on how quickly we want the reasons to shift from 1 to 0 when the reason thresholds are crossed.

\subsection{Motion Planning and Control Implementation}
\label{sectionD}
To demonstrate the generalisability and practical implementability of our framework, we integrate the human reason-based supervision framework into an AV feature that governs its behaviour. In this research, we adopt an established motion planning and control framework \cite{rahmani2023bi}. In the following, we briefly describe the underlying motion planning and control mechanism and explain how the reason-triggered replanning seamlessly fits into this structure.

\subsubsection{Motion Planning}
The motion planner in this study builds a directed graph from the vehicle's current state using pre-computed motion primitives. An A* algorithm is applied to find the optimal path by minimizing the cost function
\begin{align}\label{eq:a_star_j}
J_{\text{path}} = &w_1 \cdot J_\text{length} + w_2 \cdot J_\text{smoothness} \nonumber\\
&+ w_3 \cdot J_\text{obstacle\_clearance} + w_4 \cdot J_\text{traffic\_rule}
\end{align}

where $J_\text{length}$ is the cost related to the length of the path from the initial state to the goal state, aiming to motivate the shortest path; $J_\text{smoothness}$ is the cost related to the smoothness of the path; $J_\text{obstacle\_clearance}$ is the cost for avoiding obstacles; and $J_\text{traffic\_rule}$ is the cost aiming to avoid areas prohibited by traffic rules. The output of the planner is a reference trajectory passed to the controller for execution. We employ a modified version of A* to enhance search efficiency and applicability for our specific use case. The detailed algorithm implementation is documented in \cite{Rahmani_2025_Decentralized}. It's worth noting that the cost function has been slightly modified to fit the purpose of this study. More specifically, the weights related to the costs for obstacle clearance and traffic adherence have been separated compared to the standard implementation in \cite{Rahmani_2025_Decentralized}.

\subsubsection{Controller}
We formulate a finite-horizon optimisation problem that is solved at each time step to ensure trajectory following while respecting user-defined constraints.

The vehicle state at the time step $t$ is represented as:
\begin{equation}
\mathbf{x}(t) = [x_t, y_t, \theta_t, v_t]^T
\end{equation}
comprising position coordinates $(x_t, y_t)$, heading angle $\theta_t$, and longitudinal velocity $v_t$. The control inputs are:
\begin{equation}
\mathbf{u}(t) = [a_t, \delta_t]^T
\end{equation}
where $a_t$ denotes acceleration and $\delta_t$ the steering angle.

Vehicle dynamics are modelled using a bicycle model as follows:
\begin{equation}
    \begin{aligned}
        \dot{x} = v \cos(\theta + \beta),  
        \dot{y} = v \sin(\theta + \beta),
        \dot{\theta} = \frac{v}{L} \sin(\beta), 
        \dot{v} = a
    \end{aligned}
\end{equation}

with slip angle $\beta = \arctan(\frac{l_r}{L}\tan(\delta))$, wheelbase $L$, and rear axle distance $l_r$. We discretise this continuous model using time step $T_s$:
\begin{equation}
\mathbf{x}(t + 1) = A_d\mathbf{x}(t) + B_d\mathbf{u}(t) + \mathbf{d_d}
\end{equation}
where the discrete system matrices are:
\begin{equation}
A_d = \begin{bmatrix}
1 & 0 & T_s c_\theta & -T_s v_t s_\theta \\
0 & 1 & T_s s_\theta & T_s v_t c_\theta \\
0 & 0 & 1 & 0 \\
0 & 0 & \frac{T_s \tan(\delta_t)}{L} & 1
\end{bmatrix}
\end{equation}

with $c_\theta = \cos(\theta_t)$ and $s_\theta = \sin(\theta_t)$ for brevity. The input matrix and disturbance term are:
\begin{equation}
B_d = \begin{bmatrix}
0 & 0 \\
0 & 0 \\
T_s & 0 \\
0 & \frac{T_s v_t}{L \cos^2(\delta_t)}
\end{bmatrix}
\end{equation}

\begin{equation}
\mathbf{d_d} = \begin{bmatrix}
T_s v_t s_\theta \theta_t \\
-T_s v_t c_\theta \theta_t \\
0 \\
-\frac{T_s v_t \delta_t}{L\cos^2(\delta_t)}
\end{bmatrix}
\end{equation}

Our cost function integrates multiple objectives over prediction horizon $N$:
\begin{equation}
\begin{split}
J = \sum_{t=0}^{N-1} &(\|e^\perp_t\|^2_{Q_\perp} + \|e^{\|}_t\|^2_{Q_{\|}} + \|e_{\theta v,t}\|^2_{Q_{\theta v}}) \\
&+ \sum_{t=0}^{N-1} (\|\mathbf{u}_t\|^2_R + \|\Delta\mathbf{u}_t\|^2_{R_d}) \\
&+ \|\mathbf{x}_N - \mathbf{x}_{ref,N}\|^2_{Q_f}
\end{split}
\end{equation}

where $e^\perp_t$ and $e^{\|}_t$ represent perpendicular and parallel trajectory tracking errors, $e_{\theta v,t}$ captures orientation and velocity errors, and $\Delta\mathbf{u}_t = \mathbf{u}_{t+1} - \mathbf{u}_t$. The matrices $Q_\perp$, $Q_{\|}$, $Q_{\theta v}$, $R$, $R_d$, and $Q_f$ are weighting matrices that prioritize different aspects of performance.

The optimisation operates under constraints:
\begin{align}
\mathbf{x}_{t+1} = f(\mathbf{x}_t, \mathbf{u}_t) ; \mathbf{u}_t &\in \mathcal{U} ;
\mathbf{x}_t \in \mathcal{X}
\end{align}

where $\mathcal{U}$ defines input limitations:
\begin{align}
a_{min} \leq a_t \leq a_{max};
\delta_{min} \leq \delta_t \leq \delta_{max}
\end{align}

\subsubsection{Reason-Triggered Replanning}
At each time step, the system evaluates the reason scores for all stakeholders using the formulations provided in Eq. \ref{eq:policy_reason} to Eq. \ref{eq:vehicle_reason}.
For each stakeholder $s_i \in \mathcal{S}$, the reason score $R_{s_i}(\mathbf{x_t}, \mathbf{u_t})$ is computed. If any score falls below its corresponding threshold $\tau_{s_i}$,
\begin{equation}
\min_{s_i \in \mathcal{S}} R_{s_i}(\mathbf{x_t}, \mathbf{u_t)} < \tau_{s_i},
\end{equation}
the system immediately triggers a replanning cycle. During replanning, the current scenario is updated, and a new reference trajectory is generated and passed to the controller. This continuous evaluation ensures that the vehicle's motion remains aligned with the ethical and performance criteria of all stakeholders. In our implementation, the path finding algorithm incorporates a set of weights in its cost function (Eq.~\ref{eq:a_star_j}) to provide flexibility when replanning is needed. For instance, under normal circumstances, prohibited areas by traffic rules are treated similarly to obstacles by assigning large weights to the $J_\text{traffic\_rule}$, restraining the A* search algorithm from generating paths through those areas. When replanning is triggered due to misalignment with human reasons, prohibited areas could temporarily receive lower costs, allowing the A* algorithm to search through those areas and provide a new trajectory with a different, potentially higher, human-reasoning score. Since the replanning strategy is not within the scope of this study, we refer the readers to the implementation of our planner detailed in \cite{Rahmani_2025_Decentralized}. We would like to highlight that the proposed evaluation framework remains algorithm-agnostic and can assess trajectories generated by any motion planning approach.

\section{EXPERIMENT SETUP}\label{sec:experiment}
\begin{figure}[t] \centering \includegraphics[width=6 cm]{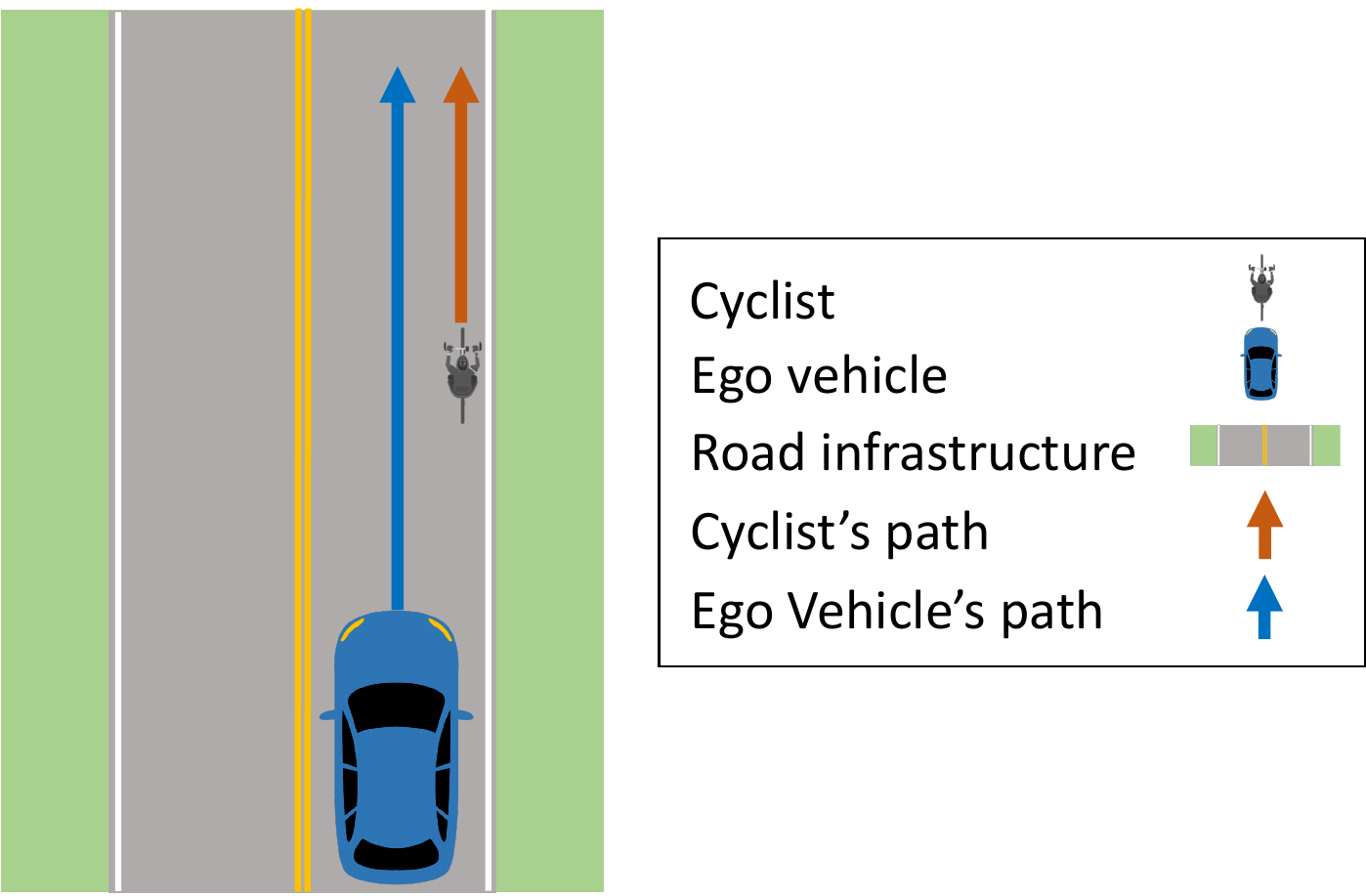} \caption{Illustration of an ego vehicle approaching a cyclist on a narrow bidirectional road, highlighting the ethical challenge in overtaking due to road constraints.} \label{fig:prob_layout} \end{figure}

To evaluate our human reasons-based supervision framework, we test it in an ethically challenging cyclist overtaking scenario, where an ego vehicle traveling in the right lane encounters a slow-moving cyclist on a narrow road (Fig. \ref{fig:prob_layout}). Safely overtaking requires the vehicle to briefly enter the left lane, which is normally reserved for oncoming traffic. This forces a trade-off between strict lane adherence and efficient, safe manoeuvring, highlighting the ethical dilemma arising from the conflicting priorities of the involved stakeholders:

\begin{itemize} 
    \item \textit{Road Policymakers} enforce traffic regulations that prohibit left-lane usage to ensure overall road safety. 
    \item \textit{Cyclists} require a safe and comfortable riding experience, which may be compromised by vehicles manoeuvring too closely. 
    \item \textit{Drivers} aim for efficient travel, potentially pressuring the system to overtake despite the inherent safety and regulatory concerns. 
\end{itemize}

The detailed definitions of the reason models for each stakeholder are provided in Eq. \ref{eq:policy_reason} to Eq. \ref{eq:vehicle_reason}. In our experiments, we focus on comparing two configurations: \begin{itemize} 
    \item \textit{Baseline Controller:} The ego vehicle operates using a standard baseline controller without the human reasons-based supervision \cite{rahmani2023bi}. 
    \item \textit{Baseline Controller with Replanner:} The baseline controller is augmented with the human reasons-based supervision framework, which triggers replanning when the vehicle's behaviour does not align with the predefined ethical thresholds. \end{itemize}

These experiments are designed to assess how integrating human reasons-based supervision framework impacts decision-making in vehicle behaviour during ethically challenging situations; specifically in the context of safely overtaking a cyclist on a bidirectional road. To calibrate our reason models, we set the threshold values summarised in Table \ref{tab:parameters} based on empirical studies of cyclist and driver behaviour \cite{oskina2023safety, hagemeister2024reported}.

\begin{table}[htbp]
\centering
\caption{Parameter Values for Reason Models}
\begin{tabular}{ll}
\hline
Parameter & Value \\
\hline
\(d_{\text{th,vru}}\) (Cyclist’s perceived too-close distance threshold) & 8 m \\
\(t_{\text{th,vru}}\) (Max time cyclist tolerates close following) & 5 s \\
\(d_{\text{th,driver}}\) (Driver’s perceived too-close distance threshold) & 12 m \\
\(t_{\text{th,driver}}\) (Max time driver tolerates close following) & 10 s \\
\(\tau_{s_{i}}\) (Reason alignment threshold for all stakeholders) & 0.7 \\
\hline
\end{tabular}
\label{tab:parameters}
\end{table}

\section{RESULTS}\label{sec:results}
\begin{figure}[h]
    \centering    \includegraphics[width=0.45\textwidth]{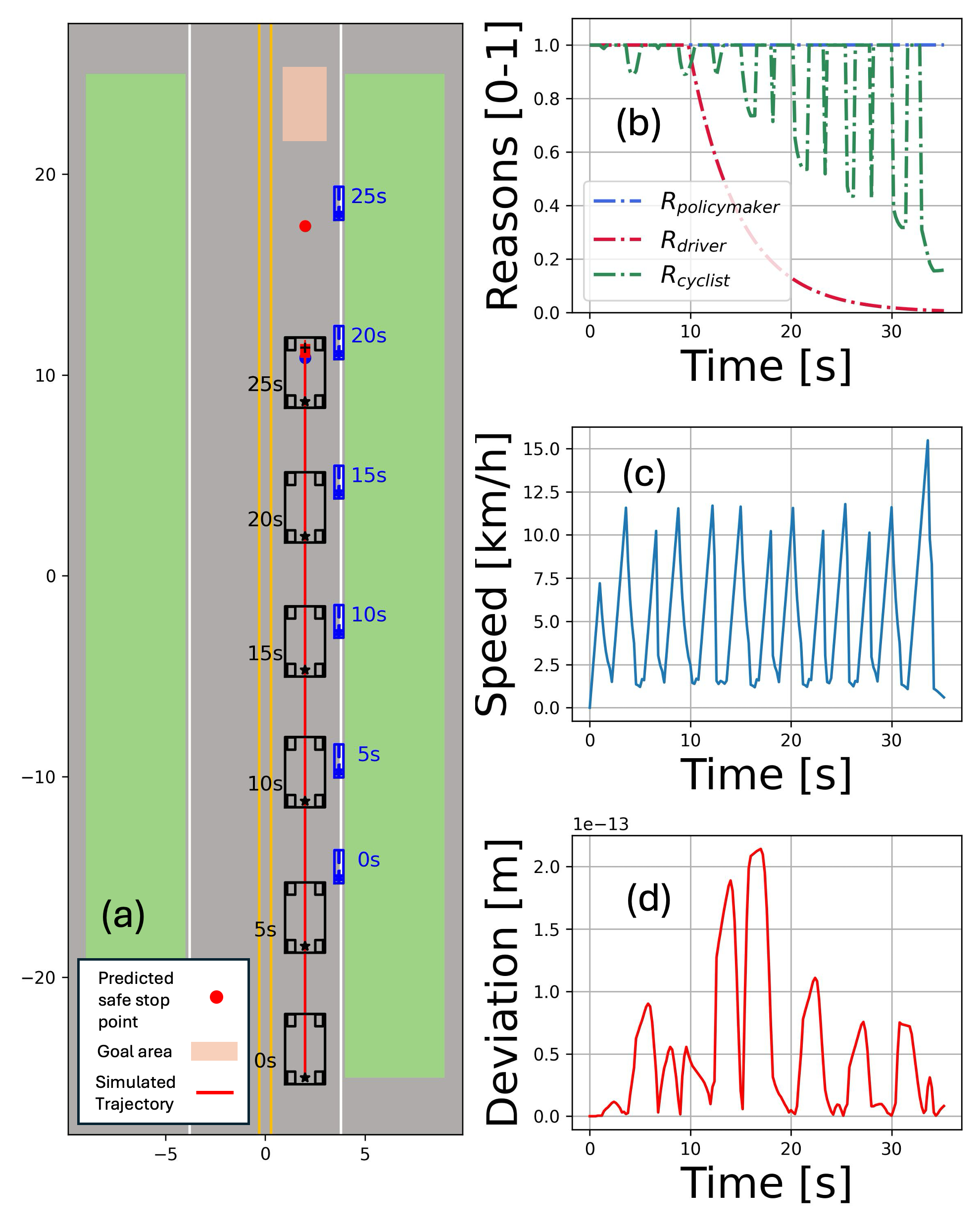}
    \caption{Results of running the model in baseline controller}
    \label{fig:scenario1}
\end{figure}

\begin{figure}[t]
    \centering    \includegraphics[width=0.45\textwidth]{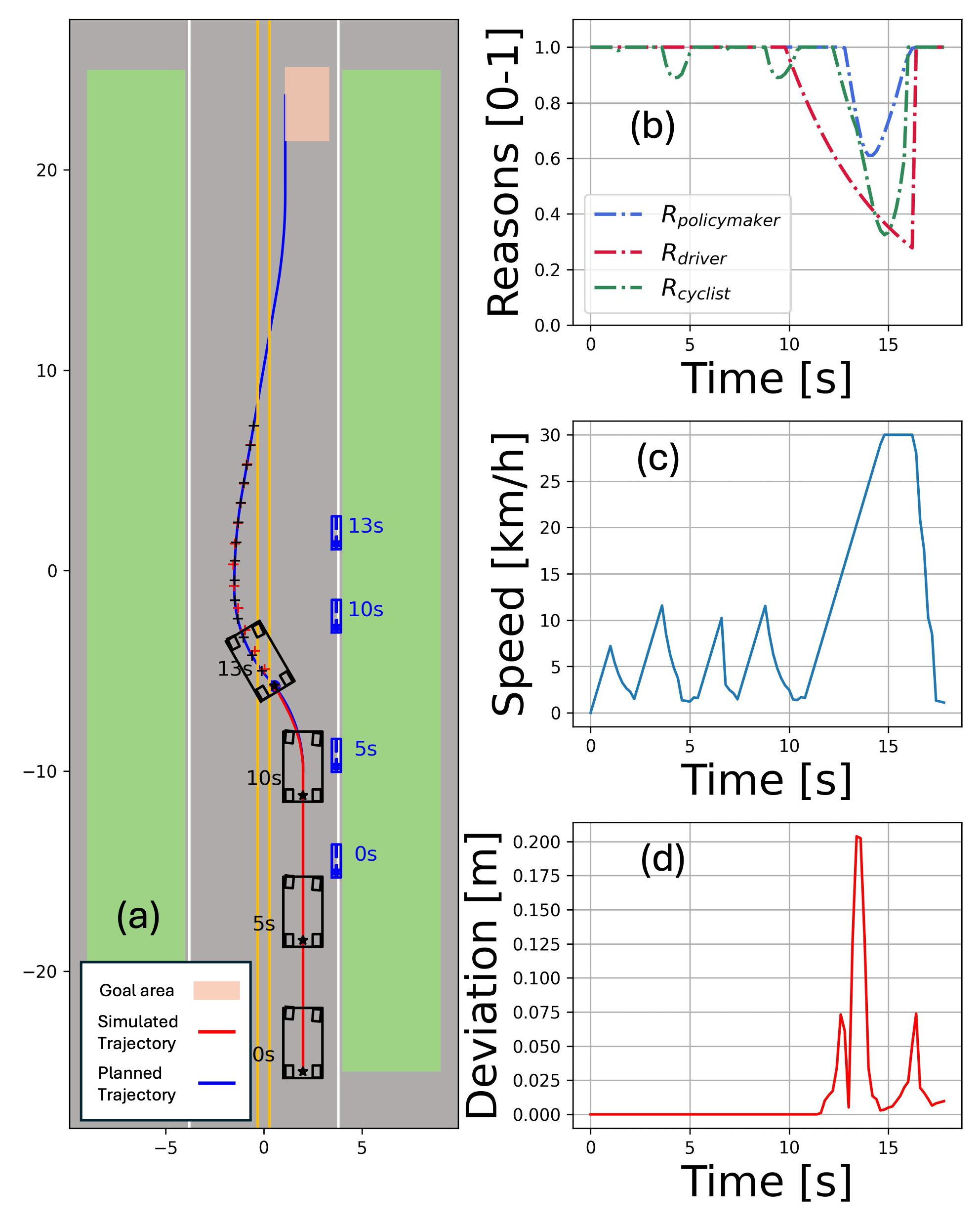}
    \caption{Results of running the model in baseline controller with replanner}
    \label{fig:scenario2}
\end{figure}

The results of validation for the controller with and without the proposed human reasons-based supervision framework are depicted in Fig.~\ref{fig:scenario1} and Fig.~\ref{fig:scenario2}, respectively. The timestamps next to the ego vehicle and the cyclist indicate their positions at that time, helping to visualize their relative movements. The results for the \textit{Baseline controller} suggest that while the system successfully handles basic path planning and collision avoidance, it does not adequately account for human reasons, particularly in terms of the driver's and cyclist's perspectives. As shown in Fig.~\ref{fig:scenario1}.a, the ego vehicle follows the cyclist and reaches the goal in 35 seconds without attempting to overtake. This behaviour demonstrates a stop-and-go dynamic, which is further illustrated in Fig.~\ref{fig:scenario1}.c, where the speed of the ego vehicle is depicted. Initially, the global planner generates a smooth path toward the goal. However, when the ego vehicle approaches the cyclist and a potential collision risk arises, the controller activates a collision avoidance strategy, reducing the ego vehicle's speed to avoid the potential collision. As the distance between the two increases, the controller allows the vehicle to accelerate and realign with the planned path.

Despite effectively tracking the planned trajectory (Fig. \ref{fig:scenario1}.d), the system's performance in aligning with human reasons decreases over time (Fig. \ref{fig:scenario1}.b). It is apparent that from the 10th second until the end of the simulation, the driver's reason score for time efficiency decreases sharply to zero by the end of the simulation. This is because the driver must remain patient to stay in the mode of following the vehicle from behind. On the other hand, the cyclist's reason score for comfort fluctuates (due to the fluctuations of the vehicle's speed and distance to the cyclist) but ultimately forms a decay pattern. Over time, the score decreases further due to the accumulation of time spent being followed by the ego vehicle at a close distance. Meanwhile, the system's performance demonstrates strong alignment with the road policymaker's reason score for regulatory compliance since the ego vehicle consistently stays in the right lane.

The \textit{Baseline controller with a replanner} allows the ego vehicle to successfully overtake the cyclist, addressing human reason priorities but at the cost of temporary regulatory compliance violations. Initially, the ego vehicle follows the cyclist from behind for the first 11 seconds, adhering to its straight path trajectory while exhibiting stop-and-go behaviour, as shown in Fig. \ref{fig:scenario2}.c. At the 11.5-second mark, the human reason-based supervision framework detects that the driver's reason score for time efficiency has fallen below its threshold of 0.7, due to the accumulation of the waiting time of the driver and the cyclist. This triggers the planner to generate a new feasible trajectory. This new path briefly crosses the bidirectional road before returning to the right lane to reach the goal. During the overtaking manoeuvre, the close proximity between the ego vehicle and the cyclist causes a temporary decrease in reason scores, and the violation of the right-lane regulation further reduces the policymaker's reason score. However, once the ego vehicle successfully overtakes the cyclist and returns to the intended lane, all reason values recover to one. In this scenario, the ego vehicle achieves the goal in just 18 seconds by overtaking the cyclist, significantly reducing the driver's waiting time. It is worth noting that the deviations on the order of centimeters in both scenarios may be attributed to the limitations and constraints of the MPC.
\section{DISCUSSION}\label{sec:discussion}
This study reveals three key contributions of the human reasons-based supervision framework: (1) the ability to detect when current system behaviour no longer aligns with the priorities of human stakeholders by monitoring reason score againts adaptive thresholds, (2) the modularity of the framework to adapt controller behaviour without majorly modifying core components like the global planner or MPC settings, and (3) the inherent explainability of decision-making processes, enabling autonomous systems to justify behavioural changes based on stakeholder reason alignment. These features are essential for building trust and acceptance in automated vehicle deployment. The experimental results yield several insights. While both the \textit{Baseline Controller} and the \textit{Baseline Controller with Replanner} successfully guide the ego vehicle to follow the planned trajectory and reach the goal area, they differ significantly in how they respond to human reasons. This leads to distinct trade-offs in performance and alignment with the priorities of human reasons.

The \textit{Baseline Controller} demonstrates a conservative approach, prioritizing the policymaker's reason for regulatory compliance. This results in strict adherence to rules, but at the cost of neglecting other human reasons priorities. While the controller achieves basic objectives like collision avoidance and goal attainment, it fails to address the driver's reason for time efficiency and the cyclist's reason for comfort.

In contrast, the \textit{Baseline Controller with Replanner} introduces a dynamic and adaptive approach through the human reasons-based supervision component of the framework. Our results show that responding to triggers can lead to decisions that better reflect a balance of human reasons. For example, the system may temporarily violate regulatory compliance (lowering the policymaker's reason score) to reduce discomfort for the cyclist or impatience from the driver. While the trigger does not resolve value conflicts, it signals when the current trajectory may no longer align with a stakeholder’s reasons. However, how the planner could systematically select among alternatives is beyond the scope of this research. Future work is needed to extend this supervision layer with decision-making mechanisms that actively weigh and decide how to respond when human reasons are in conflict.

The choice of threshold values plays a critical role in this framework. A lower threshold might delay intervention, leading to prolonged misalignment with human reasons, while a higher threshold could result in overly frequent replanning, which increases the computational cost. Additionally, the vehicle's state when the reasons falls below the threshold---such as its proximity to the cyclist, speed, or surrounding environment---can influence the feasibility of the replanned trajectory. These factors highlight the importance of carefully pick the right threshold to ensure feasibility and stability.

However, while the threshold offers an interpretable mechanism for initiating replanning, it does not capture the full nuance of how human drivers make context dependent trade-offs, such as deciding when it is safe to pass with oncoming traffic or assessing visibility in hilly terrain. Rather than prescribing the best course of action, the framework uses thresholds to signal that the current plan may no longer reflect certain stakeholder priorities. Grounding threshold values in empirical studies, and learning or tuning them from human data, is a promising direction for future work.

Nonetheless, we emphasise that our human reasons-based supervision framework, which triggers replanning based on threshold values, is not intended to compete with existing nuanced decision-making algorithms for dynamic environments, such as multi-policy decision-making (MPDM) proposed by \cite{cunningham2015mpdm, mehta2016autonomous}, but to complement them. While our framework has lower resolution than MPDM’s continuous policy evaluation, its strength lies in simplicity. It avoids the computational cost of constant replanning by activating only when a misalignment with human reasons is detected.

Future work could integrate MPDM-style approaches into our architecture, enabling motion planners that not only generate but also evaluate candidate trajectories based on human reasons rather than predefined policies. This integration could support more nuanced balancing of stakeholder priorities while preserving interpretability.

Overall, by aligning system behaviour with human priorities, the proposed framework enables more human-centric decision-making, which is essential for user trust and acceptance in real-world applications. Thanks to the framework's modularity, its integration into existing automated system architectures is straightforward. Its implementation can be extended to more complex environments, such as urban driving or multi-agent systems, where balancing multiple reasons priorities is critical. Future work could explore enhancing the framework's capabilities, such as using it not only to trigger replanning but also to identify trajectories that maximise human reasons across all agents. Testing in dynamic and unpredictable environments would further validate its robustness and scalability. 

This study underscores the importance of incorporating human reason into automated driving systems. The findings demonstrate that while strict regulatory compliance ensures safety and rule-following, mechanisms that detect misalignment with human priorities and prompt reconsideration can lead to decisions that better reflect the reasons of multiple stakeholders. This insight paves the way for future developments in automated systems that are both technically robust and socially and ethically aligned with human reasons.


\section{CONCLUSION}\label{sec:conclusion}
This study proposes a human based-reason supervision framework to support automated vehicles (AVs) to navigate routine yet ethically challenging scenarios. The framework introduces a novel approach to AV planning by evaluating whether the vehicle's behaviour aligns with human reason and triggering a replan assignment if misalignment is detected. The key contributions demonstrated through this work are: (1) A detection mechanism for identifying misalignment between AV behaviour and stakeholder reasons based on reason score thresholds; (2) Modular integration into the AV control architecture without modification of the core planner or motion controller; (3) Explainability through the use of stakeholder reason scores, enabling interpretable justifications for behavioural changes. These features enable AVs to align with human reasons in real time, ensuring more human-centric decision-making. 

\section{ACKNOWLEDGMENTS}
We declare that OpenAI ChatGPT was used in the writing process to enhance the style and conciseness of the text originally written by the authors.  This work was funded by the Indonesia Endowment Fund for Education (LPDP) under Grant 0006552/TRA/D/19/lpdp2021.

\bibliographystyle{IEEEtran}

\end{document}